\documentclass[conference]{IEEEtran}
\IEEEoverridecommandlockouts
\usepackage{cite}
\usepackage{amsmath,amssymb,amsfonts}
\usepackage{algorithmic}
\usepackage[linesnumbered,ruled,vlined]{algorithm2e}
\usepackage{graphicx}
\usepackage{textcomp}
\usepackage{xcolor}
\usepackage{mathtools}
\usepackage{microtype}
\usepackage{booktabs} % for professional tables
\usepackage{caption}
\usepackage{subcaption}
\usepackage{float}

\def\BibTeX{{\rm B\kern-.05em{\sc i\kern-.025em b}\kern-.08em
    T\kern-.1667em\lower.7ex\hbox{E}\kern-.125emX}}
\begin{document}
\thispagestyle{plain}
\pagestyle{plain}
\title{AlphaRouter: Quantum Circuit Routing with Reinforcement Learning and Tree
Search}

\author{
\IEEEauthorblockN{Wei Tang\textsuperscript{\textdagger}}
\IEEEauthorblockA{AWS Quantum Technologies \\
New York, NY, USA \\
weittang@amazon.com}
\and
\IEEEauthorblockN{Yiheng Duan\textsuperscript{\textdagger}}
\IEEEauthorblockA{AWS Quantum Technologies \\
Seattle, WA, USA \\
yiheng@amazon.com}
\and
\IEEEauthorblockN{Yaroslav Kharkov}
\IEEEauthorblockA{AWS Quantum Technologies \\
New York, NY, USA \\
ykharkov@amazon.com}\\
\and  
\IEEEauthorblockN{Rasool Fakoor}
\IEEEauthorblockA{Amazon Web Services \\
Santa Clara, CA, USA \\
fakoor@amazon.com}
\and
\IEEEauthorblockN{Eric Kessler}
\IEEEauthorblockA{AWS Quantum Technologies \\
New York, NY, USA \\
erikessl@amazon.com}
\and
\IEEEauthorblockN{Yunong Shi}
\IEEEauthorblockA{AWS Quantum Technologies \\
New York, NY, USA \\
shiyunon@amazon.com}
}

\maketitle
\begingroup\renewcommand\thefootnote{\textdagger}
\footnotetext{Wei Tang and Yiheng Duan contributed equally to this work.}
\endgroup
\begin{abstract}
Quantum computers have the potential to outperform classical computers in important tasks such as optimization and number factoring. They are characterized by limited connectivity, which necessitates the routing of their computational bits, known as qubits, to specific locations during program execution to carry out quantum operations.  Traditionally, the NP-hard optimization problem of minimizing the routing overhead has been addressed through sub-optimal rule-based routing techniques with inherent human biases embedded within the cost function design. This paper introduces a solution that integrates Monte Carlo Tree Search (MCTS) with Reinforcement Learning (RL). 
Our RL-based router, called AlphaRouter, outperforms the current state-of-the-art routing methods and generates quantum programs with up to $20\%$ less routing overhead, thus significantly enhancing the overall efficiency and feasibility of quantum computing.
\end{abstract}

% \begin{IEEEkeywords}
% component, formatting, style, styling, insert
% \end{IEEEkeywords}

\section{Introduction}
% What is the problem being studied?
Quantum computing~\cite{nielsen2001quantum, feynman1982simulating} is an emerging computational paradigm, which has a potential to transform many industries, including optimization~\cite{Farhi2014AQA}, machine learning \cite{Schuld2021}, quantum chemistry and material science~\cite{AspuruGuzik2005SimulatedQC, Bauer2020QuantumAF}. A quantum computer comprises quantum bits (qubits) that share quantum edges with each other. Quantum computers perform computation by applying quantum operations (gates) to qubits, akin to classical Boolean circuits. Such quantum gates operate on pairs of qubits, creating correlation in their states. This phenomenon, called entanglement, constitutes the underlying potential advantage of quantum computing.

However, quantum computers face a key challenge in applying quantum gates. Due to physical constraints, quantum gates can only be applied to adjacent pairs of qubits on a quantum computer. In general, quantum computers do not have all-to-all connectivity. As a result, transforming high-level abstract quantum programs (logical circuits) into hardware executable quantum circuits (physical circuits) that satisfy the connectivity constraints is necessary to run quantum programs.

% Why should people care about it?
Quantum routers hence become a necessary component to tackle the challenge. To accommodate the sparse qubit connectivity in quantum computers, routers must move pairs of remote logical qubits to adjacent physical qubits, so that they can be executed on a quantum computer. To generate physical circuits that comply with specific topologies, routers utilize specialized quantum operations known as SWAP gates. These gates are applied to adjacent pairs of physical qubits to exchange the positions of the corresponding pair of logical qubits, effectively swapping their locations.

Unfortunately, SWAP insertion is very costly. Quantum operations are inherently non-perfect and error-prone~\cite{Franklin2004ChallengesIR}. SWAPs introduce significantly more error to quantum programs, destroys accurate quantum data and makes quantum program outputs wrong. As a result, optimizing routers to minimize their added SWAP overhead is crucial for advancing quantum computing efficiency~\cite{murali2019noise}.

% What are the current solutions and their limitations?
Minimizing SWAP gates in quantum programs is crucial but challenging, as it parallels the token swapping problem~\cite{yamanaka2015swapping}, recognized as at least NP-hard~\cite{cowtan2019qubit}. Currently, there are two primary approaches to the circuit routing problem: satisfiability solver-based routing algorithms~\cite{Lin2023ScalableOL} and heuristics-based routing algorithms~\cite{cowtan2019qubit, anis2021qiskit, Kharkov2022ArlineBA}. However, solver-based algorithms achieve optimal SWAP numbers but suffer from exponential time complexity with more qubits and quantum gates. On the other hand, heuristic routers generally produce a suboptimal solution with a high SWAP overhead. In contrast, machine learning methods for circuit routing problem allow to balance the solution quality and the runtime.

% What is our key idea?
%Instead, quantum circuit routing can be formulated as a sequential decision making process by employing a RL-based methods. In addition to the recent advancements in RL that have led to impressive performance in simulable digital environments like video games~\cite{silver2016mastering}, RL frameworks have also been widely adopted in many complex applications such as autonomous driving~\cite{kiran2021deep}, datacenter cooling~\cite{lazic2018data}, navigation of stratospheric balloons, and more. Firstly, a RL-based router adapts to different benchmarks and quantum computer topologies simply through re-training or fine-tuning, eliminating the need for manual algorithm redesign. Secondly, RL agent models, particularly those utilizing transformers~\cite{vaswani2017attention}, are adept at handling longer input sequences. This capability enhances their potential to achieve global optimization results, thereby significantly improving the efficiency of RL-based routing over traditional heuristic methods.
Instead, quantum circuit routing can be formulated as a sequential decision-making process by employing RL-based methods. Given the recent impressive results of Deep RL~\cite{mnih2013playing, silver2016mastering, silver2018general, fakoorp3o, fakoor2019metaqlearning, kiran2021deep, lazic2018data, bellemare2020autonomous,fakoor2021continuous, bugdetYao23, alphatensor2022}, adapting Deep RL frameworks to this problem offers several advantages. Firstly, an RL-based router can adapt to different benchmarks and quantum computer topologies through re-training or fine-tuning, eliminating the need for manual algorithm redesign. Secondly, RL agent models, particularly those utilizing transformers~\cite{vaswani2017attention}, are adept at handling longer input sequences. This capability enhances their potential to achieve better optimization results, thereby significantly improving the efficiency of RL-based routing over traditional heuristic methods.

% How do we solve the problem with this key idea?
We developed an RL-based quantum circuit router, called AlphaRouter. AlphaRouter is inspired by AlphaZero~\cite{silver2016mastering}, where it combines MCTS~\cite{Coulom2006EfficientSA, Kocsis2006BanditBM} and RL~\cite{sutton2018reinforcement}. MCTS excels in exploring a vast space of possible actions and outcomes, enhancing decision-making by balancing exploration and exploitation. Meanwhile, RL contributes by learning from the outcomes of these explorations, continuously improving strategy over time. This synergy allows for more sophisticated and globally optimized quantum compilation solutions. Our contributions hence include:

% What are our contributions and why they are important. Summarize the key results.
\begin{enumerate}
    \item \textbf{SWAP reduction}: To the best of our knowledge, AlphaRouter is the first to integrate MCTS and RL to optimize for extended circuit sequences, in contrast against traditional methods that focus on single-layer local search. This enables AlphaRouter to reduce the number of SWAP gates by $10-20\%$ over state-of-the-art routers.
    \item \textbf{Unseen Benchmark}: AlphaRouter demonstrates robust  performance and generalization ability across benchmarks, achieving consistent compilation results even for previously unseen benchmarks.
    \item \textbf{Scalability} AlphaRouter exhibits scalable compilation efficiency, evidenced by a $15\%$ reduction in the linear scaling coefficient for the number of SWAPs as benchmark sizes increase, while retaining a low inference time. 
\end{enumerate}

\section{Background}
\subsection{Quantum Circuit Routing}
\begin{figure*}[t]
    \centering
    \begin{subfigure}{0.2\textwidth}
        \centering
        \includegraphics[width=\linewidth]{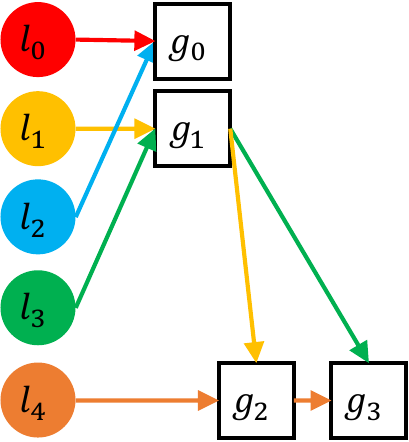}
        \caption{Input logical circuit as a Directed Acyclic Graph (DAG) with five logical qubits $l_i,\forall{i}\in\{0,\ldots,4\}$ and four quantum gates $g_j,\forall{j}\in\{0,\ldots,3\}$.}
        \label{fig:remaining_circuit_1}
    \end{subfigure}
    \hfill
    \begin{subfigure}{0.3\textwidth}
        \centering
        \includegraphics[width=\linewidth]{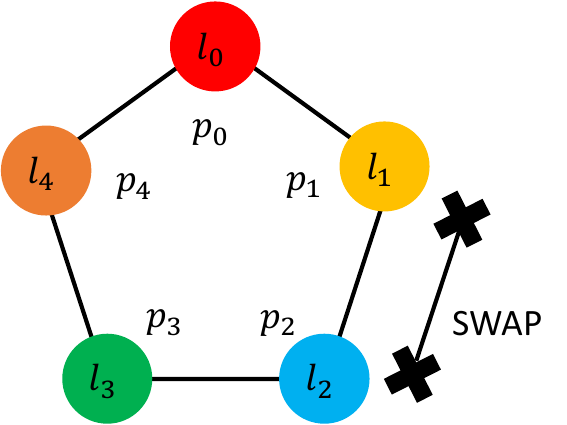}
        \caption{A SWAP between physical qubits $p_1$ and $p_2$ exchanges $l_1$ and $l_2$.}
        \label{fig:qpu_1}
    \end{subfigure}
    \hfill
    \begin{subfigure}{0.3\textwidth}
        \centering
        \includegraphics[width=\linewidth]{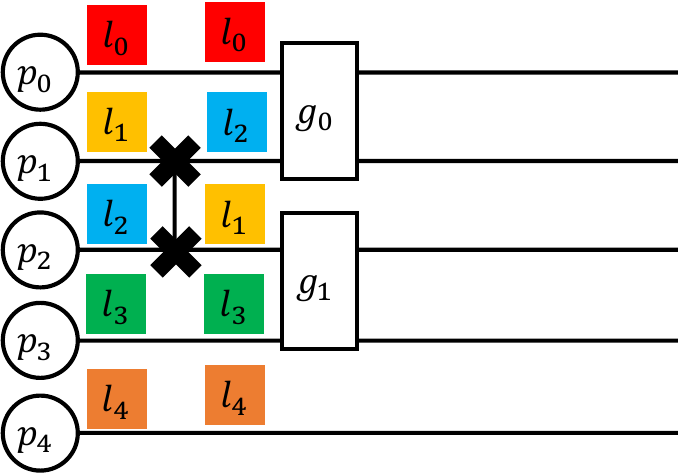}
        \caption{$g_0$ can be scheduled as $l_0$ and $l_2$ are adjacent. $g_1$ can be scheduled as $l_1$ and $l_3$ are adjacent.}
        \label{fig:output_1}
    \end{subfigure}
    \vfill
    \begin{subfigure}{0.2\textwidth}
        \centering
        \includegraphics[width=\linewidth]{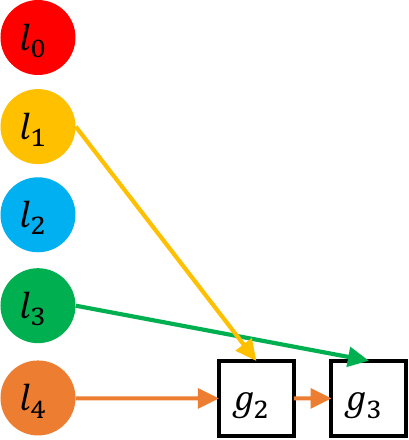}
        \caption{Remaining logical circuit after the first SWAP.}
        \label{fig:remaining_circuit_2}
    \end{subfigure}
    \hfill
    \begin{subfigure}{0.3\textwidth}
        \centering
        \includegraphics[width=\linewidth]{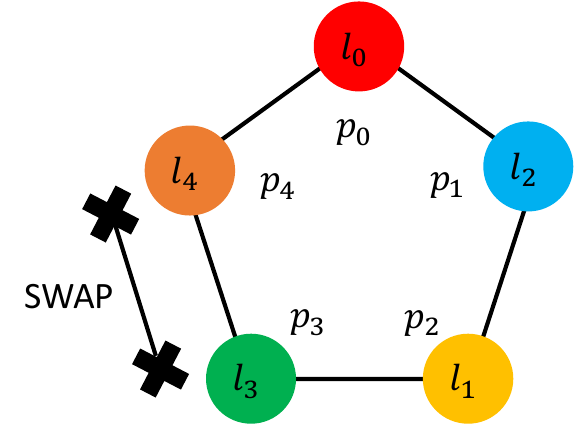}
        \caption{A SWAP between physical qubits $p_3$ and $p_4$ exchanges $l_3$ and $l_4$.}
        \label{fig:qpu_2}
    \end{subfigure}
    \hfill
    \begin{subfigure}{0.3\textwidth}
        \centering
        \includegraphics[width=\linewidth]{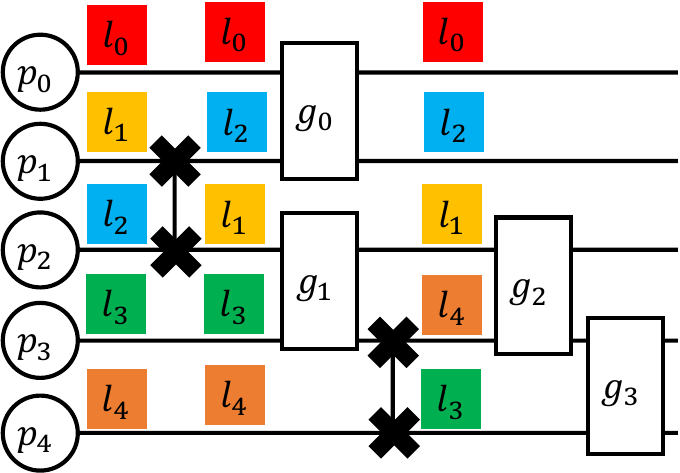}
        \caption{$g_2$ can be scheduled as $l_1$ and $l_4$ are adjacent. $g_3$ can be scheduled as $l_3$ and $l_4$ are adjacent and its preceding gate $g_2$ has also been scheduled.}
        \label{fig:output_2}
    \end{subfigure}
    \caption{Example of routing a logical circuit to a quantum computer using two SWAPs. Figures~\ref{fig:remaining_circuit_1},~\ref{fig:remaining_circuit_2} show the logical circuits. Colors represent the flow of qubits through the gates. Figures~\ref{fig:qpu_1},~\ref{fig:qpu_2} show the topology and qubit mapping of a ring quantum computer. $p_i,\forall i\in\{0,\ldots,4\}$ represent physical qubits. $l_i,\forall i\in\{0,\ldots,4\}$ represent the logical qubit mapped to a physical qubit. The black edges represent the connections. Figures~\ref{fig:output_1},~\ref{fig:output_2} represent the routed physical circuit output. The gates are scheduled on their target logical qubits and topology compliant.}
    \label{fig:compilation_example}
\end{figure*}

Figure~\ref{fig:remaining_circuit_1} shows an example $5$-qubit logical circuit visualized as a Directed Acyclic Graph (DAG). Logical quantum gates could apply to any pair of logical qubits. Each logical qubit goes through one or more quantum gates to evolve its quantum state. This is analogues to the classical computing process of applying operations to data stored in registers.

Figure~\ref{fig:qpu_1} shows the topology and qubit mapping of a ring quantum computer with a trivial initial mapping, i.e. logical qubit $l_i$ is mapped to physical qubit $p_i$, $\forall i\in\{0,\ldots,4\}$. Given the logical circuit and the quantum computer mapping, $g_0$ cannot be scheduled since the corresponding physical qubits of $l_0$ and $l_2$ -- $p_0$ and $p_2$ -- are not adjacent on the quantum computer. $g_1$ cannot be scheduled for the same reason. In addition, $g_2$ and $g_3$ must wait for $g_1$ to execute first because of data dependency. Instead, inserting a SWAP between $p_1$ and $p_2$ brings pairs of logical qubits $(l_{0}, l_2)$ and $(l_{1}, l_3)$ adjacent on the quantum computer. Figure~\ref{fig:output_1} hence shows that both gates $g_{0}$ and $g_1$ can now be scheduled into the output physical circuit.

Figure~\ref{fig:remaining_circuit_2} shows the remaining logical circuit after the first SWAP. Since the logical qubit targets of $g_2$ and $g_3$ are not adjacent on the quantum computer, they require an additional SWAP between $p_3$ and $p_4$, as shown in Figure~\ref{fig:qpu_2}. Consequently, Figure~\ref{fig:output_2} shows that a second SWAP between $p_3$ and $p_4$ brings logical qubits $(l_{1}, l_4)$ adjacent on the quantum computer, thus enabling $g_2$. With its dependence on $g_2$ fulfilled, $g_3$ is also scheduled, hence completing the routing process.

Routing strategically inserts SWAPs to reposition logical qubits on a quantum computer so that they become adjacent to execute the required logical gates. A SWAP could be applied to any physical edge on a quantum computer, exchanging the position of the two logical qubits. This process is guided by the quantum computer topology, the current qubit mapping and the remaining sequence of logical gates. Routing carries out a sequential decision making process to apply a SWAP gate and schedule all possible logical gates, given the current state. The routing process terminates once all the logical gates from the input circuit are successfully scheduled into the output physical circuit.

Existing quantum circuit routing methods~\cite{murali2019noise,li2019tackling,cowtan2019qubit}  adopt local search that only considers the first few gates from the front layer of the circuit, leading to suboptimal routing quality.
\subsection{Reinforcement Learning}
\begin{figure}[t]
\vskip 0.2in
\begin{center}
\centerline{\includegraphics[width=.4\textwidth]{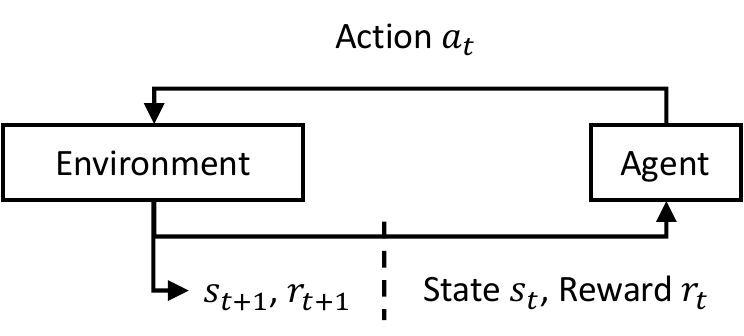}}
\caption{A standard RL framework. An agent interacts with an environment to generate and learn from its experiences.}
\label{fig:rl_framework}
\end{center}
\vskip -0.2in
\end{figure}

In Reinforcement learning (RL), an agent learns to make decisions by interacting with an environment~\cite{sutton2018reinforcement}. It involves the agent taking actions in the environment and receiving feedback in the form of rewards. Markov Decision Process (MDP)~\cite{PutermanMDP1994} is a mathematical framework commonly used to model and formalize RL problems. An MDP is defined by a tuple $(\mathcal{S}, \mathcal{A}, T, r, \mu_0, \gamma)$, where $\mathcal{S}$ denotes the state space and $\mathcal{A}$ is the action space, the function $T : \mathcal{S} \times \mathcal{A} \times \mathcal{S} \rightarrow \mathbb{R}_+$ encodes the transition probabilities of the MDP, $\mu_0$ denotes the initial state distribution, $r(s,a)$ is the instantaneous reward obtained by taking action $a \in A$ in state $s \in \mathcal{S}$, and $\gamma \in [0, 1]$ is a discount factor for future rewards. The goal of RL is to find a policy, which is a mapping from states to actions $f : \mathcal{S} \rightarrow \mathcal{A}$ , that maximizes the cumulative expected reward over time.

Figure~\ref{fig:rl_framework} visualizes the standard RL framework. An agent, usually in the form of a neural network, interacts with an environment by taking actions $a_t$ based on the current state $s_t$. Each action then leads to a new state $s_{t+1}$ and produces a reward $r_t$, guiding the agent to learn optimal behaviors through trial and error to maximize cumulative rewards.

Given the sequential nature of quantum circuit routing, RL can be an effective tool for finding optimal routing. Section~\ref{sec:problem_statement} details the formalization of quantum circuit routing, framing it as an MDP planning problem, and demonstrates how RL can efficiently and adaptively optimize the process.
\subsection{Monte Carlo Tree Search}\label{sec:mcts}
\begin{figure}[t]
    \centering
    \begin{subfigure}{0.23\textwidth}
        \centering
        \includegraphics[width=\linewidth]{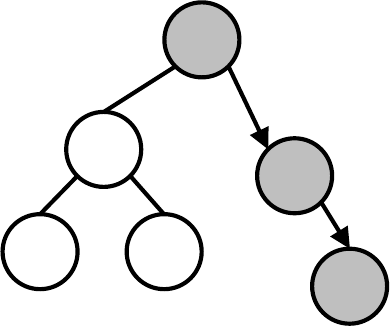}
        \caption{Selection. Greedy traversal of the existing tree from the root state to choose the leaf state with maximum UCB scores.}
        \label{fig:mcts_selection}
    \end{subfigure}
    \hfill
    \begin{subfigure}{0.23\textwidth}
        \centering
        \includegraphics[width=\linewidth]{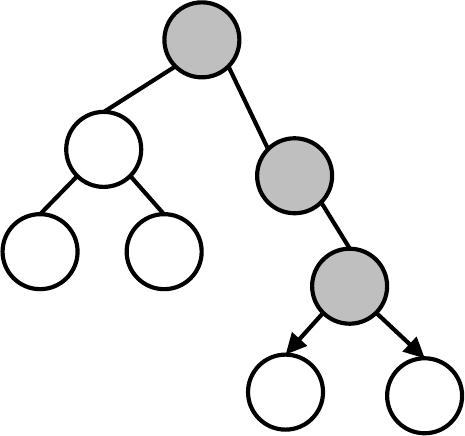}
        \caption{Expansion. Add child states by applying possible actions.}
        \label{fig:mcts_expansion}
    \end{subfigure}
    \hfill
    \begin{subfigure}{0.23\textwidth}
        \centering
        \includegraphics[width=\linewidth]{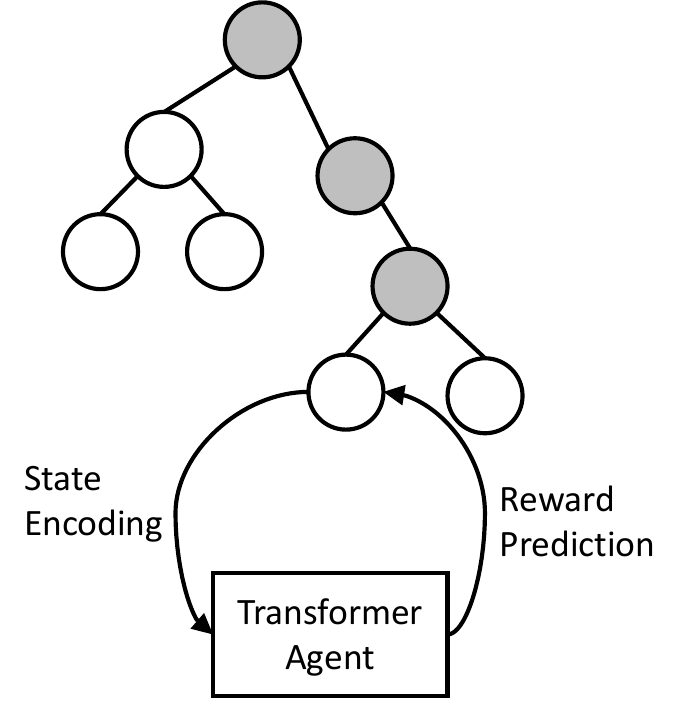}
        \caption{Simulation. Agent predicts the additional reward from the leaf state to a terminal state.}
        \label{fig:mcts_simulation}
    \end{subfigure}
    \hfill
    \begin{subfigure}{0.23\textwidth}
        \centering
        \includegraphics[width=\linewidth]{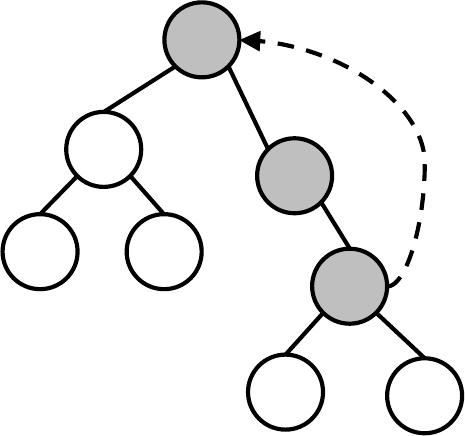}
        \caption{Backpropagation. Update the statistics of the branch of nodes involved in the current iteration.}
        \label{fig:mcts_back_propagation}
    \end{subfigure}
    \caption{MCTS with a transformer agent. The process repeats the four stages for a pre-determined max number of iterations. Circles in the tree represent states. Lines connecting the states represent actions.}
    \label{fig:mcts}
\end{figure}

MCTS~\cite{Coulom2006EfficientSA} is a search algorithm that progressively builds a search tree by exploring and evaluating potential moves, while focusing on exploring the high value regions. MCTS does not require supervision, but generates values for states iteratively.

Figure~\ref{fig:mcts} shows the four stages in an MCTS cycle. An MCTS search tree consists of nodes corresponding to states $s$. In addition, each node stores statistics such as the number of visits $N(s)$ and the Monte Carlo valule estimation $Q(s)$. Throughout the iterative cycles of these stages, the values associated with each state in the tree are continuously updated.

The Selection stage greedily traverses the existing search tree from the root state till reaching a non-terminal child state. Greedy selections are based on maximising the Upper Confidence Bound (UCB)~\cite{kocsis2006bandit} score in Equation~\ref{eq:ucb}.
\begin{equation}
UCB(s_t\to s_{t+1})=\frac{Q(s_{t+1})}{N(s_{t+1})}+c\times\sqrt{\frac{\log N(s_t)}{N(s_{t+1})}}\label{eq:ucb}
\end{equation}
where $s_t$ indicates a parent node, $s_{t+1}$ represents a child node to $s_t$, $Q(s)$ represents the current Monte Carlo value of a state $s$, $N(s)$ represents the number of visits to a state $s$, and $c$ represents a hyperparameter tradeoff coefficient. UCB scores hence favor children states less visited and with higher Monte Carlo values.

The Expansion stage appends one or more new child states to the selected leaf state to expand the tree by applying each possible action. This represents exploring a new action from the leaf state.

The Simulation stage conducts a simulated roll-out from the leaf state by predicting the additional reward to a potential terminal state. Vanilla versions of MCTS usually follow a random policy to reach a terminal state. For example, when searching next moves for the game of Go, a vanilla MCTS implementation randomly plays stones to the current board until victory or loss. However, quantum circuit compilation is a decision process with an infinite horizon. This means that randomly inserting SWAP gates to the circuit does not guarantee compilation completion. Instead, AlphaRouter obtains a prediction of the value of the leaf state by feeding the leaf state encoding to a transformer-based agent, described in section~\ref{sec:agent}.

The Backpropagation stage propagates the real rewards gathered from the simulation back to the root state. The statistics for each node within the selected branch, including metrics like value and visit count, are updated to reflect the results obtained from the simulation stage. This process ensures that the information at each node accurately represents the accumulated knowledge so far.
\section{Problem Statement}\label{sec:problem_statement}
Finding the minimum number of SWAPs to route a logical quantum circuit to a quantum computer can be modeled as an MDP. AlphaRouter focuses training a routing agent for a given quantum computer topology, i.e. with fixed number of qubits and connectivity. A quantum computer is modeled as a set of edges $E\equiv \{e:p_i, p_j\}$, where $P\equiv\{p_j\}$ represents the physical qubits. Similarly, $L\equiv\{l_j\}$ represents the logical qubits in an input quantum circuit.

The state space for circuit routing $s_t$ comprises of the remaining logical circuit to be routed and the current qubit mapping. The state space is infinite as there are infinitely many logical cirucits. A logical quantum circuit is expressed as logical gates $G_t\equiv \{g:l_{g,1}, l_{g,2}\}$, where $l_{g,1}$ and $l_{g,2}$ represent the two logical qubits of a gate $g$. The qubit mapping is a bijective mapping between physical and logical qubits $M_t:P\longleftrightarrow L$.

The action space is limited as there are $|E|$ possible SWAPs to insert, one for every quantum computer edge. A SWAP action $A_t$ on edge $\{p_i, p_j\}$ exchanges the qubit mapping from $p_i,p_j\leftrightarrow l_i,l_j$ to $p_i,p_j\leftrightarrow l_j,l_i$ and updates the mapping $M_t$ to $M_{t+1}$.

In addition, gates are sequentially checked for topology compliance and pending dependence on other gates. $g\in G_t$ satisfying the following conditions are scheduled and removed from $G_t$: (i) $l_{g,1},l_{g,2}$ are adjacent on the quantum computer after the SWAP, i.e. $(M_{t+1}(l_{g,1}),M_{t+1}(l_{g,2}))\in E$ and (ii) $l_{g,1},l_{g,2}$ have no other pending gates in $G_t$ preceding $g$ that had not been removed in the current time step.

The circuit routing problem is hence framed as: Route an input logical quantum circuit $G_0$ to a quantum computer backend $E$ with an initial mapping $M_0$. Find the shortest sequence of actions $\{s_t, A_t, s_{t+1}\}_{t=0}^{T-1}$ such that $|G_T|=0$.
\section{Framework}

AlphaRouter combines both RL and MCTS to autonomously train a RL agent to predict the best placement of SWAPs given a logical quantum circuit and a quantum computer.

\subsection{Training}\label{sec:training}
\begin{figure*}[t]
\vskip 0.2in
\begin{center}
\centerline{\includegraphics[width=.7\textwidth]{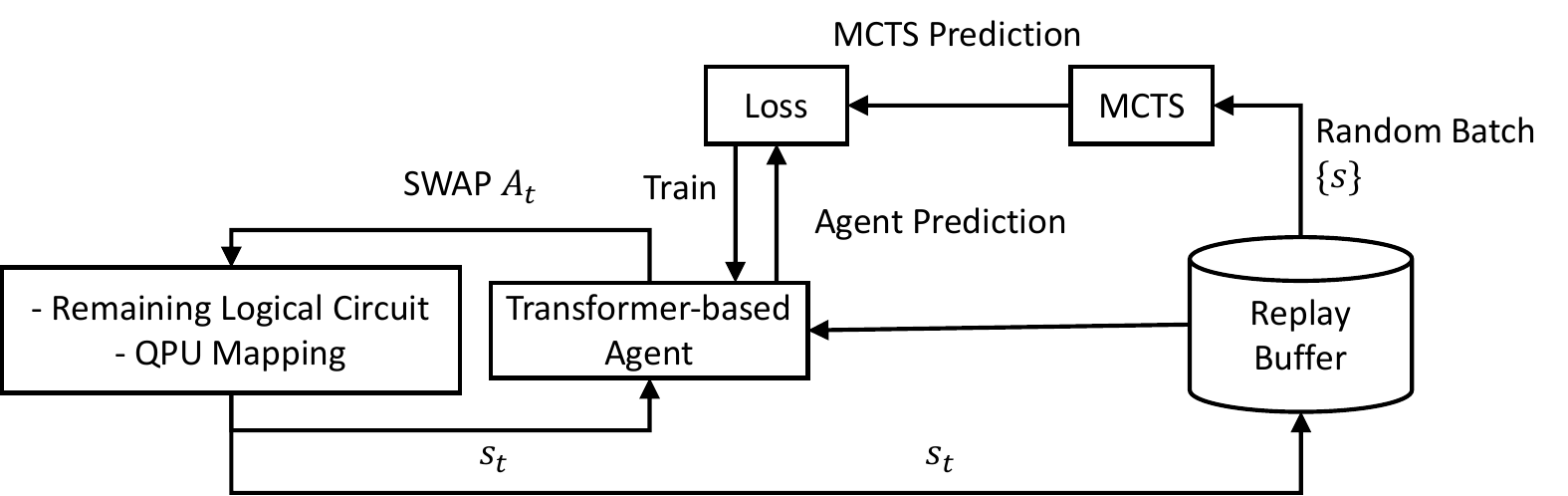}}
\caption{AlphaRouter: MCTS + RL training framework.}
\label{fig:training_framework}
\end{center}
\vskip -0.2in
\end{figure*}

Our training approach, inspired by AlphaZero, enables autonomous label generation and more efficient exploration for RL via MCTS. Figure~\ref{fig:training_framework} visualizes the overall training framework.
The environment comprises both the remaining logical circuit to be routed and the current qubit mapping to produce a state observation $s_t$. Section~\ref{sec:agent} details the state encoding. In addition, we limit the remaining logical circuit lookahead to be up to $48$ logical gates. A transformer-based agent predicts the next SWAP action $A_t$ by feeding the state $s_t$ through its network as shown in Figure~\ref{fig:agent_network}. A chosen SWAP action is then applied to advance the environment state.

A replay buffer stores the state experiences $s_t$ and passes a randomly sampled experience batch $\{s\}$ to MCTS. MCTS undergoes its value update process in Section~\ref{sec:mcts} for the sampled batch for $200$ iterations (rollouts). Rewards $r_t$ are defined to be the number of additional gates scheduled by applying action $A_t$, minus a penalty of incurring one extra SWAP:
\begin{equation}
r_t=|G_{t}|-|G_{t+1}|-1\label{eq:reward}
\end{equation}
Note that $r_t=-1$ if a SWAP fails to schedule any logical gates at time step $t$.

MCTS then outputs its action prediction for the sampled states $s'\in\{s\}$.
\begin{equation}
    a^*_{MCTS}=arg\max_a{Q^{MCTS}(f(s',a))}
\end{equation}
where $f$ is a state transition function $s_{t+1}=f(s_t,a)$, and $Q^{MCTS}(s)$ is the Monte Carlo value of a state $s$.

In addition, the agent outputs its action probability prediction by forwarding $s'\in\{s\}$ through its transformer network. The two state value predictions are then compared to compute a cross entropy loss:
\begin{equation}
    l_1(s')=-\sum_a\delta(a^*_{MCTS},a)\times\log(Q^\pi(s',a))\label{eq:cross_entropy}
\end{equation}
where $\delta(a,a')$ is a delta function between two actions $a$ and $a'$. $Q^\pi(s,a)$ is the transformer network action probability prediction for a given state $s$ and action $a$.

The two state value predictions are also compared to compute a square loss:
\begin{equation}
    l_2(s')=(Q^{MCTS}(s')-Q^\pi(s'))^2\label{eq:mse}
\end{equation}
where $Q^\pi(s)$ is the transformer state value prediction for a state $s$.

Finally, an average loss combines both the cross entropy loss in Equation~\ref{eq:cross_entropy} and the square loss in Equation~\ref{eq:mse} to update the transformer network.
\begin{equation}
    L=\frac{l_1+\alpha l_2}{|E|}\label{eq:loss}
\end{equation}
where $|E|$ is the size of the action space. $\alpha$ is a hyperparameter tradeoff coefficient. $\alpha$ balances the two losses $l_1$ and $l_2$. $l_1$ is the cross entropy loss between transformer network action probability prediction and MCTS action prediction. $l_2$ is the squared loss between MCTS and transformer SWAP count prediction for a given state. The tradeoff coefficient $\alpha$ is needed because larger benchmarks will require more SWAPs to finish compilation. As a result, $l_2$ naturally becomes larger and dominant for larger benchmarks. $\alpha$ is empirically determined based on the size of input benchmarks such that $l_1$ and $l_2$ are roughly of the same scale.

One of the key advantages of MCTS is its highly parallelizable nature, primarily due to the structure of its data and the relative independence of node expansions. This aspect of MCTS makes it well-suited for distribution across multiple compute nodes. In our implementation, we leverage this parallelizability by distributing the MCTS process across $16$ $ml.c5.18xlarge$ CPU instances on AWS SageMaker. Each instance independently handles a portion of the tree search and node expansion, significantly enhancing the overall efficiency and speed of the search process. By dividing the workload in this manner, we can concurrently execute multiple tree searches and updates, accelerating the MCTS value update and making the most of the available computational resources.
\subsection{Routing as Inference}
\begin{figure}[t]
\vskip 0.2in
\begin{center}
\centerline{\includegraphics[width=.4\textwidth]{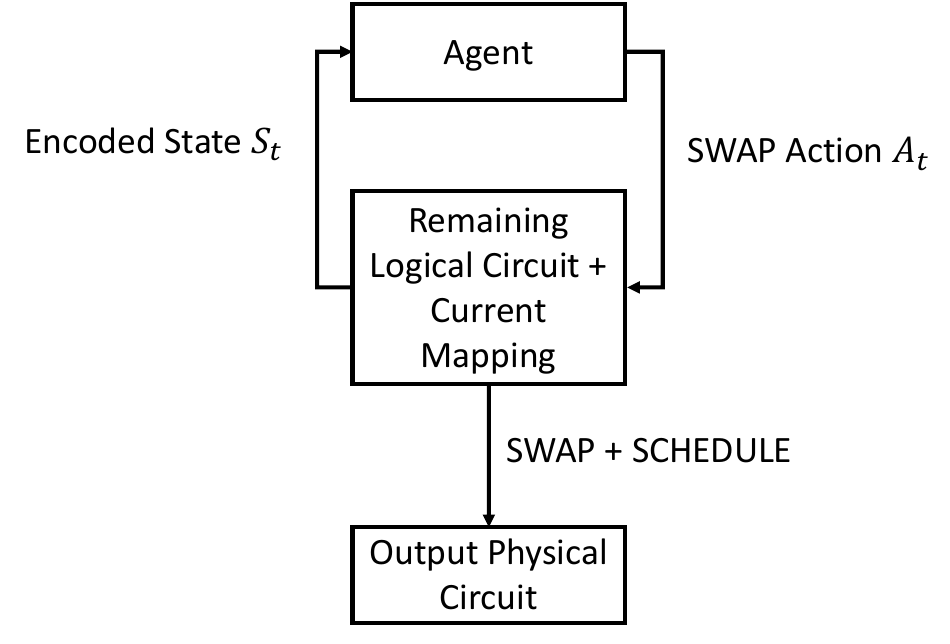}}
\caption{AlphaRouter compiles quantum circuits with its trained agent via model inference without MCTS.}
\label{fig:inference_framework}
\end{center}
\vskip -0.2in
\end{figure}
After completing its training, the RL agent employs a standard iterative inference process to perform circuit routing. Figure~\ref{fig:inference_framework} illustrates AlphaRouter's use of a trained agent for routing via model inference. The RL agent receives an encoded state $s_t$ comprising the remaining logical circuit and current qubit mapping. Based on this, it then predicts and applies the next SWAP action $a_t$ to adjust the qubit mapping. Any logical gates that now satisfy the topology are greedily scheduled. The iteration terminates when the input logical circuit has been fully routed.

In the standard AlphaZero~\cite{silver2016mastering} implementation, MCTS is utilized alongside the trained agent during the inference phase for additional refinement of predictions. However, this integrated approach tends to be slower compared to using the trained agent alone.

AlphaRouter, having demonstrated performance improvements over existing methods, enables us to omit the use of MCTS during inference to achieve a much faster runtime. Our experiments show that excluding MCTS from the inference process significantly improves the runtime without compromising the routing performance. This enhancement in runtime is crucial for quantum computing, particularly because numerous quantum algorithms, such as variational quantum algorithms, necessitate the execution of thousands of distinct circuits to tackle benchmark problems. For context, the operational timeframe for executing a quantum circuit on a quantum processor is typically within the microsecond range. Consequently, the time taken for compilation can quickly emerge as a significant bottleneck in the standard workflow of quantum computing.
\subsection{Agent Network}\label{sec:agent}
\begin{figure*}[t]
\vskip 0.2in
\begin{center}
\centerline{\includegraphics[width=.7\textwidth]{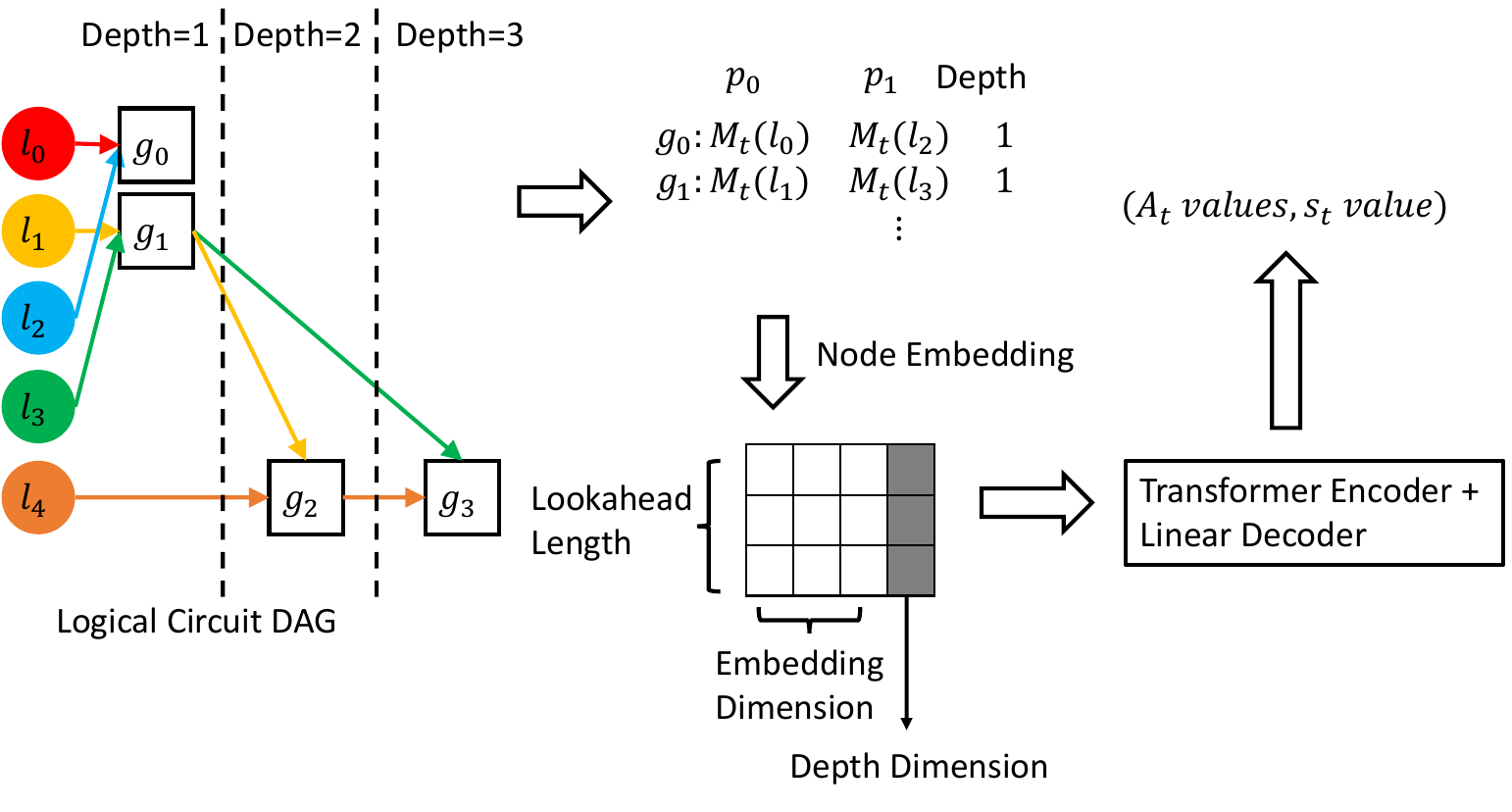}}
\caption{Encoding the logical circuit $G_t$ and the current qubit mapping $M_t$.}
\label{fig:agent_network}
\end{center}
\vskip -0.2in
\end{figure*}

\begin{figure}[t]
\vskip 0.2in
\begin{center}
\centerline{\includegraphics[width=.3\textwidth]{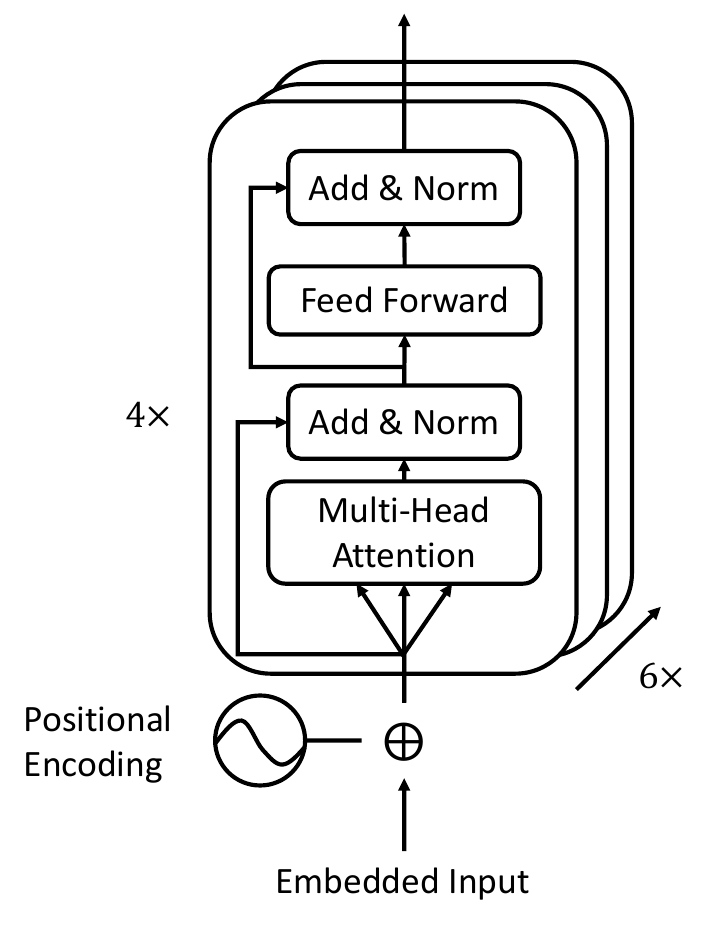}}
\caption{AlphaRouter uses a transformer-based neural network as its agent. The model comprises $4$ multi-head attention layers, each layer has $6$ attention heads.}
\label{fig:transformer}
\end{center}
\vskip -0.2in
\end{figure}

Figure~\ref{fig:agent_network} depicts the state encoding, beginning with the logical circuit DAG. In this DAG, each quantum gate is represented by its two qubits and a depth based on its topological order. For example, $g_0$ has physical qubit targets of $M(l_0)$ and $M(l_2)$ based on the mapping $M$. In addition, since $g_0$ does not depend on any other gates, it has a depth of $1$. $g_{2}$ has direct dependency on $g_1$ and hence have depth of $2$. Eventually, $g_3$ has depth of $3$ as it depends on $g_2$.

AlphaRouter observes the first few logical gates to limit the lookahead depth with a max number set at $48$. Each gate vector then embeds its two physical qubit targets with a trainable lookup table. Each gate vector is subsequently represented as the mean of the two qubit embeddings, together with an extra dimension of depth. Overall, the node embedding process produces a feature matrix of shape $\textrm{Lookahead Length}\times(\textrm{Embedding Dimension}+1)$.

Finally, this matrix is fed into a transformer-based actor-critic~\cite{konda1999actor} network for encoding and decoding, yielding predictions including both SWAP action values and a state value. Figure~\ref{fig:transformer} shows the transformer network architecture. 
AlphaRouter employs a $4$-layer, $6$-head transformer encoder architecture, starting with the embedded input shown in Figure~\ref{fig:agent_network}.

The embedded input is first combined with positional encoding to retain sequence information. The Transformer architecture lacks inherent positional information in its structure because it processes input sequences as sets of elements without considering the order of these elements. To address this, Transformers often incorporate positional encodings~\cite{vaswani2017attention} to provide the model with information about the position of each element in the sequence.

Positional encoding involves alternating sine and cosine functions that depend on the position of a quantum gate in the input sequence and the embedding dimension. This positional information is crucial for quantum circuit routing as the order of quantum gates carries physical meaning. Equations~\ref{eq:positional_encoding} shows the sinusoidal wave functions.
\begin{eqnarray}
    PE(pos,2i)&=&\sin{(pos/10000^{2i/d})}\nonumber\\
    PE(pos,2i+1)&=&\cos{(pos/10000^{2i/d})}\label{eq:positional_encoding}
\end{eqnarray}

Each layer comprises two main components: multi-head attention and a feed-forward network, both followed by an ``Add \& Norm'' step for residual connections and layer normalization. The multi-head attention, with $6$ attention heads, allows the model to focus on different parts of the input sequence simultaneously, enhancing its ability to capture complex dependencies. This is because self-attention enables each gate to interact with every other gate in the circuit, assessing their relationships and the overall sequence. Such an approach captures not only the individual qubit connections of each gate but also how gates influence one another across the circuit. This nuanced understanding of both local and global gate interactions is crucial for identifying optimal routing paths, as it mirrors the inherently interconnected nature of quantum computations, where the placement and order of gates significantly impact the best routing sequence. This structure repeats across all $4$ layers, enabling the encoder to process and transform the input into a high-level, contextually enriched representation. A final linear layer serves as the decoder and outputs the action and state value predictions.

Table~\ref{tab:transofrmer_params} summarizes AlphaRouter's Tranformer model architecture and hyperparameters.
\begin{table}[h]
    \centering
    \begin{tabular}{c|c|c}
        Number of attention layers & 4 \\
        Number of attention heads & 6 \\
        Node embedding dimension & 20 \\
        Total number of parameters &  2.2M \\
        Optimizer &  Adam \\
        Learning rate & 0.1\\
        Learning rate decay & 0.8 \\
        UCT coefficient $c$ & $\sqrt{2}$\\
        Batch Size & $32$
    \end{tabular}
    \caption{Transformer model parameters and training configuration.}
    \label{tab:transofrmer_params}
\end{table}
\subsection{Overall Algorithm}
\begin{algorithm}
   \caption{MCTS+RL Training}
   \label{alg:alpha_router}
\begin{algorithmic}
   \STATE {\bfseries Input:} Number of circuits $N$. Logical quantum circuits $\{G_{0,i}\}$. Initial qubit mapping $\{M_{0,i}\}$. Quantum computer topology $E$. Feature extraction function $\mathcal{O}$. State transition function $f$. Transformer network $\pi$.
   \STATE Initialize $unfinishedIndices=\{1,\ldots,N\}$.
   \STATE Initialize $t=0$.
   \STATE Initialize replay buffer $B=empty$.
   \FOR{each episode}
    \FOR{$i\in unfinishedIndices$}
        \STATE State $s_{t,i}=\mathcal{O}(G_{t,i},M_{t,i})$.
        \STATE Sample $A_{t,i}$ based on $Q^\pi(s_{t,i},a)$.
        \STATE $s_{t+1,i}=f(s_{t,i},A_{t,i})$.
        \STATE Add $s_t$ to $B$.
        \IF{$s$ is a terminal state}
            \STATE remove $i$ from $unfinishedIndices$.
        \ENDIF
    \ENDFOR
    \IF{$|B|>320$}
        \STATE Random sample experience batch $\{s\}$.
        \FOR{Each state $s'\in\{s\}$}
            \STATE Run MCTS for $200$ rollouts, outputs the MCTS action prediction $a^*_{MCTS}=arg\max_a{Q^{MCTS}(f(s',a))}$.
            \STATE Compute the action value predictions by the transformer agent $Q^\pi(s',a)$.
            \STATE Compute cross entropy loss according to Equation~\ref{eq:cross_entropy}.
            \STATE Compute squared error according to Equation~\ref{eq:mse}.
        \ENDFOR
        \STATE Compute the average batch loss according to Equation~\ref{eq:loss}.
        \STATE Update the transformer parameter with the Adam optimizer.
    \ENDIF
   \ENDFOR
\end{algorithmic}
\end{algorithm}

Algorithm~\ref{alg:alpha_router} shows the overall training algorithm. The algorithm box uses the same notations as Section~\ref{sec:problem_statement}. AlphaRouter uses the Adam optimizer~\cite{kingma2014adam} to train its transformer network.
\section{Experiment Settings}
\subsection{Quantum Computer Backends}
We train and test the performance of AlphaRouter on various quantum computers include (i) $Tokyo$, an IBM quantum computer with $12$ qubits arranged in a $3\times4$ grid; (ii) $OQC$, an Oxford Quantum Circuits quantum computer with $8$ qubits arranged in a circle; and (iii) $Guadalupe$, an IBM quantum computer with $16$ qubits in a heavy hexagon topology~\cite{chamberland2020topological}.

The training was performed using AWS SageMaker cloud service. The training runs for $100$ episodes on $Random$ benchmarks for all quantum computer topologies.

\subsection{Initial Mapping}
We employ two distinct strategies for initial qubit mapping: (i) trivial and (ii) random. The trivial initial mapping simply assigns the logical qubits in a numerical order to the physical qubits, i.e. $p_i = l_i$. On the other hand, random initial mapping introduces variability by randomly assigning qubits.
AlphaRouter is designed to solve only the circuit routing problem for a given initial mapping. In the future it is viable to extend our approach to simultaneously solve both routing and initial mapping problems using the same RL model. One possible approach is to consider the initial mapping as a sequence of virtual SWAPs in the beginning of the routing.

\subsection{Benchmarks}\label{sec:benchmarks}
The benchmarks listed below are among the most commonly used quantum circuit types. We sample random circuit instances from each circuit class and compute the average SWAP count.
\begin{enumerate}
    \item $Regular$: Quantum Approximate Optimization Algorithm (QAOA) solves the maximum independent set problem for random $3$-regular graphs~\cite{saleem2020approaches}.
    \item $Erdos$: The same algorithm as $Regular$ but for random Erdos-Renyi graphs.
    \item $QFT$: Quantum Fourier Transform~\cite{cooley1965algorithm} is a common subroutine in many quantum algorithms that promise speedup over classical algorithms, including the Shor's factoring algorithm.
    \item $QV$: Quantum Volume circuits~\cite{cross2019validating} quantify the largest random circuit of equal width and depth that a quantum computer successfully implements.
    \item $GHZ$: Generates the entangled Greenberger–Horne-Zeilinger state~\cite{greenberger1989going}.
    \item $BV$. Bernstein Vazirani algorithm solves the hidden string problem more efficiently than classical algorithms do~\cite{bernstein1997quantum}.
    \item $HS$. Hidden Shift algorithm~\cite{van2006quantum}.
    \item $Random$. Random quantum circuits consisting of uniformly sampled single and two-qubit gates.
\end{enumerate}
Our benchmarks encompass a diverse array of circuit structures, ensuring a thorough and extensive assessment of our RL-based quantum circuit router's performance across various circuit complexities.

\subsection{Baseline Routers}
We use various Qiskit routers as the baseline comparisons. These are some of the most commonly used heuristics routers. Specifically, $Basic$ is a greedy strategy that finds the shortest path to implement quantum gates sequentially. $Stochastic$ is a randomized algorithm that performs multiple trials for SWAP insertion in each circuit layer. $SABRE$~\cite{li2019tackling} is the state-of-the-art heuristics router. It performs a bi-directional search and uses a custom objective function, that takes into account SWAPs both in the front layer and the last layer in the circuit.

% \subsection{Metrics}
% The primary metric for comparing the performance of different quantum circuit compilers is the number of SWAP gates used. This metric is crucial as it directly correlates with the efficiency and effectiveness of the compiler in optimizing quantum circuits. A lower count of SWAP gates indicates a more efficient compilation, reducing the overall operational complexity and potential error rates in the quantum computation process.
\section{Results}
We designed 4 experiments to test a specific aspect of the RL compiler, while keeping the other factors constant.

\subsection{Generalizing to Unseen Benchmarks}
\begin{figure}[t]
    \centering
    \begin{subfigure}{0.4\textwidth}
        \centering
        \includegraphics[width=\textwidth]{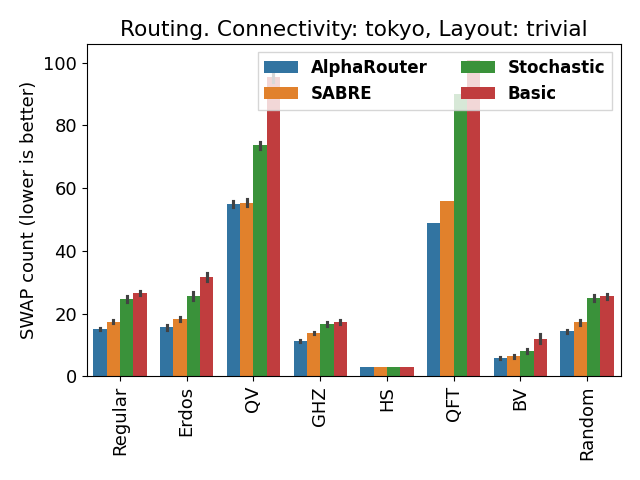}
        \caption{Trivial mapping. The error bars show the standard deviation from $20$ circuits for each benchmark type. AlphaRouter reduces the number of SWAPs by $\sim 10-15 \%$ compared to the state-of-the-art SABRE router.}
        \label{fig:trivial_mapping}
    \end{subfigure}
    \begin{subfigure}{0.4\textwidth}
        \centering
        \includegraphics[width=\textwidth]{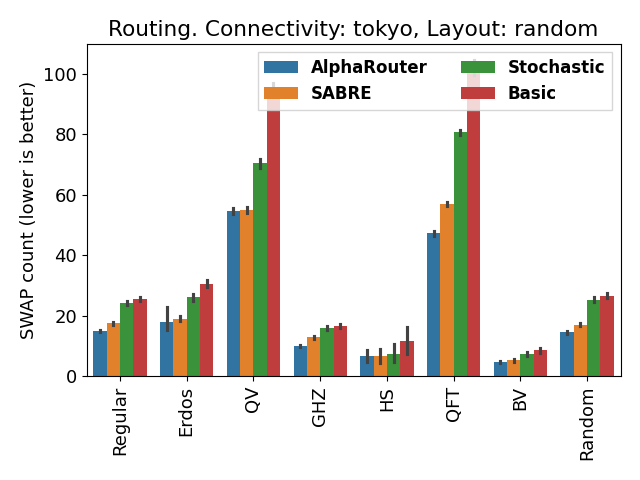}
        \caption{The same random initial mapping applies to both AlphaRouter and baseline routers. The error bars show the standard deviation from $5$ random initial mappings.}
        \label{fig:random_mapping}
    \end{subfigure}
    \caption{Average number of SWAPs by training AlphaRouter on $Regular$ and testing it on various benchmarks. AlphaRouter outperforms baseline routers for the majority of the benchmarks.}
    \label{fig:benchmark_generalization}
\end{figure}

AlphaRouter was trained on the $Regular$ benchmark and tested on a variety of benchmarks listed in Section \ref{sec:benchmarks}.
Figure~\ref{fig:benchmark_generalization} compares the total number of SWAPs by AlphaRouter and Qiskit baseline routers.
On average, the RL agent outperforms the baseline compilers for the majority of the benchmarks.
Specifically, RL reduces the SWAP counts by $10-20\%$ over the best baseline compiler in SABRE even for previously un-seen benchmarks.

\subsection{Scaling to Larger Benchmarks}
\begin{figure}[t]
\vskip 0.2in
\begin{center}
\centerline{\includegraphics[width=.9\columnwidth]{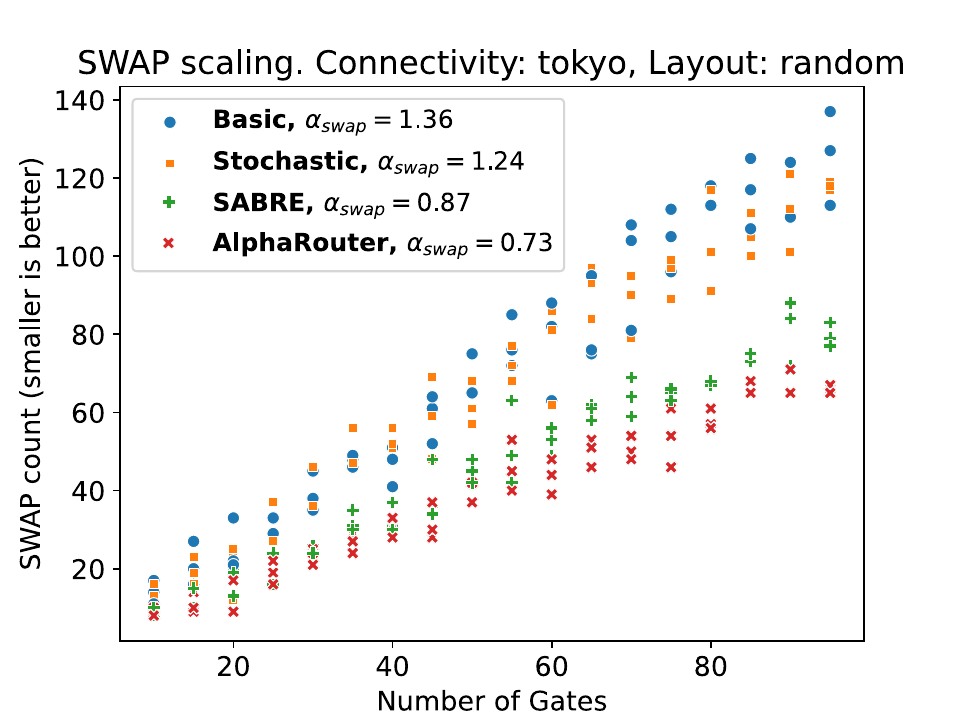}}
\caption{Routing $Random$ benchmark to the Tokyo quantum computer and random initial mapping for increasing number of gates. AlphaRouter demonstrates $\sim15\%$ smaller scaling coefficient $\alpha_{swap}$ compared to SABRE.}
\label{fig:scaling_swaps}
\end{center}
\vskip -0.2in
\end{figure}

Given a fixed benchmark type and a quantum computer topology, the number of SWAPs is expected to increase linearly with the length of the benchmark. However, this linear growth varies across different routers, indicating their routing efficiency. We route $Random$ benchmarks with an increasing number of gates on the Tokyo quantum computer using random initial mapping. Figure~\ref{fig:scaling_swaps} shows that AlphaRouter has better scaling (lower linear coefficient $\alpha$ calculated by linear regression), compared to the baseline Qiskit routers by approximately $15\%$.

\subsection{Adapting to Different Quantum Computers}
\begin{figure}[t]
\vskip 0.2in
\begin{center}
\centerline{\includegraphics[width=\columnwidth]{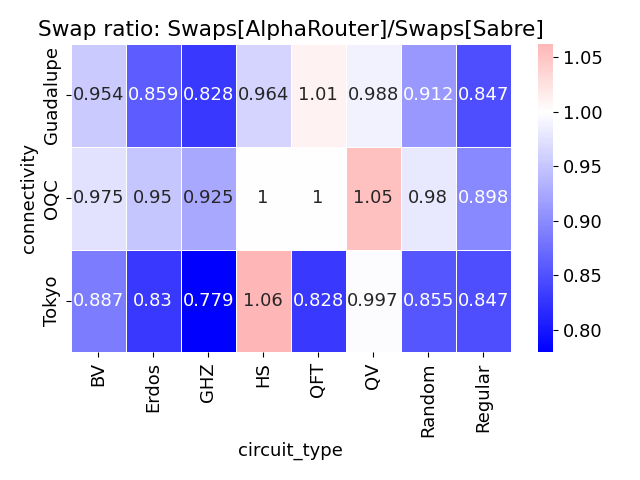}}
\caption{Average swap ratios of AlphaRouter versus SABRE across different benchmarks and with random initial mapping. Lower is better. AlphaRouter adapts to different quantum computers, outperforming or matching SABRE.}
\label{fig:heatmap}
\end{center}
\vskip -0.2in
\end{figure}

Quantum circuit routers must be able to adapt to different quantum computers. We train AlphaRouter on IBM Tokyo ($12q$), IBM Guadalupe ($16q$), and OQC ($8q$) quantum computers with $Random$ benchmark and test its performance on all benchmarks. Figure~\ref{fig:heatmap} shows the ratio of SWAPs of AlphaRouter versus SABRE. AlphaRouter excels on Tokyo and Guadalupe quantum computers, achieving $10-20\%$ SWAP reduction for most benchmarks. Its performance comes close to SABRE on the OQC quantum computer. This is because OQC is a smaller quantum computer and has less room for improvement by RL.

\subsection{Improving Initial Mapping}
\begin{figure}[t]
\vskip 0.2in
\begin{center}
\centerline{\includegraphics[width=\columnwidth]{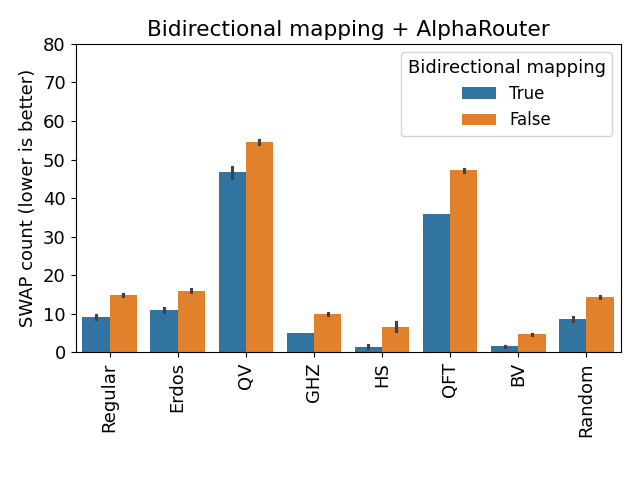}}
\caption{AlphaRouter with and without bidirectional mapping to improve its initial mapping. Integrating the technique reduces $40\%$ SWAPs on average on Tokyo quantum computer.}
\label{fig:bidirectional_mapping}
\end{center}
\vskip -0.2in
\end{figure}
The SABRE router utilizes a bidirectional mapping technique to optimize initial qubit mapping. It starts with a forward pass to establish an initial mapping of logical to physical qubits, followed by a backward pass that refines this mapping using the complete circuit layout~\cite{li2019tackling}.

While AlphaRouter only solves SWAPs, it could integrate the bidirectional mapping technique to improve its routing overhead. Experiments in Figure~\ref{fig:bidirectional_mapping} compare the AlphaRouter SWAP counts with and without the bidirectional mapping pass. Notably, incorporating bidirectional initial mapping reduces $40\%$ SWAPs on average across all benchmarks for the Tokyo quantum computer.
\section{Related Work}

Graph routing problems are well known problem in computer science and operations research~\cite{Lenstra1981ComplexityOV}. Graph routing problems can be solved by using provable algorithms with exponential runtime in the worst case, heuristics based on mixed integer programming, or by employing a combination of planning and deep reinforcement learning methods \cite{Bengio2018MachineLF, Bogyrbayeva2022LearningTS, Nazari2018ReinforcementLF, Mai2021PacketRW, Costa2021Learning2H, DNNforNetworkRouting2019}. Routing problems are inherently classical and are not specific to quantum computing.

It is widely recognized that even state-of-the-art quantum routers suffer from a significant optimality gap in SWAP count~\cite{Tan2020OptimalitySO}. \cite{pozzi2022using} have implemented a Q-learning based RL approach for quantum circuit routing, utilizing a standard setup comprising an environment and an agent. \cite{sinha2022qubit} adopted a structure combining MCTS and RL, but their optimization focuses solely on a single layer of gates. In contrast, \cite{zhou2022quantum} employed MCTS without integrating RL, offering a different perspective on quantum circuit optimization. It is important to note, that \cite{pozzi2022using, sinha2022qubit} focused on optimization of the circuit depth which defined the RL reward function, rather than optimizing the SWAP count overhead. Minimizing SWAP count is the most important metric for the efficient utilization of quantum computers rather than the gate depth. 
Each of these works contributes uniquely to the evolving landscape of quantum circuit compilation methodologies. In addition, several heuristic-based compilers~\cite{li2019tackling, cowtan2019qubit, nannicini2022optimal} perform local optimizations but lack the scalability to perform global searches.

In addition, quantum compilation also involves other directions beyond improving routing. One direction of research designs RL-based gate optimization algorithms, based on local mathematical identities between gates and local rewrite rules to simplify the input circuits before routing~\cite{fosel2021quantum, Quetschlich2022CompilerOF}.
In addition, RL has applied to continuous time control of quantum hardware, improving the accuracy of quantum operations~\cite{Yao2020NoiseRobustEQ, Bukov2017ReinforcementLI, Metz2022SelfcorrectingQM}. 
Furthermore, combining quantum and classical computation resources to perform hybrid computing has become an emerging area of quantum computing research~\cite{tang2021cutqc}.
These research cover different stages of the quantum compilation stack. 
This paper addresses the ubiquitous problem of circuit routing, which is necessary for quantum computers with a limited connectivity.
% Another stream of work is dedicated to direct synthesis of unitary gates from a fixed set of elementary gates~\cite{He2021VariationalQC, Chen2022EfficientAP}, although these methods are not scalable beyond a few qubits.
\section{Conclusion}

In this paper, we incorporate RL and MCTS to develop a quantum circuit router, distinctively free from arbitrary heuristic routing rules. Our approach not only exhibited $10-20\%$ SWAP reduction over existing routers but also demonstrated remarkable abilities in benchmark generalization and scalability. Our pioneering work in integrating MCTS with RL for quantum circuit routing highlights the transformative potential of combining RL and MCTS for complex optimization tasks, advancing the development and practicality of quantum computing.

\section{Acknowledgements}
We would like to thank Lihong Li for the insightful discussions throughout the project.
% \input{text/appendix}

% \section*{Acknowledgment}
\bibliography{references}
\bibliographystyle{plain}

\end{document}